\documentclass[aps,prd,reprint,twocolumn,10pt,eqsecnum,showpacs,nofootinbib,amsfonts,amssymb,amsmath,longbibliography,superscriptaddress,floatfix]{revtex4-1}
\usepackage[colorlinks=true,allcolors=blue, pdfusetitle]{hyperref}
\usepackage[all]{hypcap}
\usepackage{bm}
\usepackage{mathrsfs}
\usepackage[dvipsnames]{xcolor}
\usepackage{mathtools}
\usepackage{graphicx}
\usepackage{orcidlink}
\usepackage{xspace}
\usepackage{tensor}

\usepackage[T1]{fontenc}
\usepackage{lmodern}
\usepackage[utf8]{inputenc}
\usepackage{microtype}

\DeclareMathAlphabet{\mathbfsf}{\encodingdefault}{\sfdefault}{bx}{sl}

\usepackage[normalem]{ulem}

\newcommand{\Cornell}{\affiliation{Cornell Center for Astrophysics and Planetary Science, Cornell University, Ithaca, New York 14853, USA}}
\newcommand{\Columbia}{\affiliation{Columbia University, Department of Astronomy, 550 West 120th Street, New York, NY 10027, USA}}
\newcommand{\CCA}{\affiliation{Center for Computational Astrophysics, Flatiron Institute, 162 Fifth Avenue, New York, NY 10010, USA}}
\newcommand{\StonyBrook}{\affiliation{Department of Physics and Astronomy, Stony Brook University, Stony Brook, NY 11794, USA}}

\usepackage[commandprefix=ifneeded,commentmarkup=uwave,deletedmarkup=xout]{changes}
\definechangesauthor[name={Will Farr}, color=cyan]{WF}
\definechangesauthor[name={Max Isi}, color=magenta]{MI}
\definechangesauthor[name={Keefe Mitman}, color=orange]{KM}

\newcommand{\OFourRates}{180\xspace}
\newcommand{\OFiveRates}{500\xspace}

\IfFileExists{results_macros.tex}{%

\providecommand{\Nmemoryevents}{}
\renewcommand{\Nmemoryevents}{258\xspace}

\providecommand{\NForecastCrossing}{}
\renewcommand{\NForecastCrossing}{2,000\xspace}

\providecommand{\muLambdaconstraint}{}
\renewcommand{\muLambdaconstraint}{$0.29^{+2.75}_{-2.63}$\xspace}
\providecommand{\sigmaLambdaconstraint}{}
\renewcommand{\sigmaLambdaconstraint}{$2.70^{+3.03}_{-1.91}$\xspace}

\providecommand{\Aconstraint}{}
\renewcommand{\Aconstraint}{$0.26^{+4.09}_{-4.08}$\xspace}

}{}

\graphicspath{{../figures/}}

\newcommand{\figMemoryPosteriors}{%
	\begin{figure*}[t]
		\begin{center}
			\includegraphics[width=\textwidth
			]{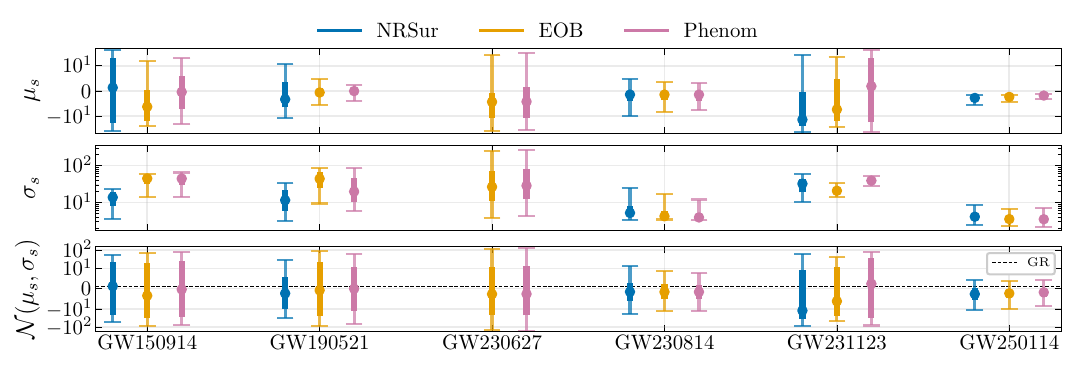}
		\end{center}
		\caption{Values of the (\emph{top}) memory mean $\mu_{s}$ (Eq.~\eqref{eq:mean}), (\emph{middle}) memory standard deviation $\sigma_{s}$ (Eq.~\eqref{eq:stddev}), and (\emph{bottom}) a draw from the $\mathcal{N}(\mu_{s},\sigma_{s})$ Gaussian for the waveform models that are available for a few events. Blue markers represent samples obtained by the \texttt{NRSur7dq4} model~\cite{Varma:2019csw}, while yellow and pink markers represent those obtained by the \texttt{SEOBNRv(4/5)PHM}~\cite{Bohe:2016gbl,Ossokine:2020kjp,Pompili:2023tna,Ramos-Buades:2023ehm} and \texttt{IMRPhenomX(PHM/PHM-SpinTaylor/O4a)} models~\cite{Pratten:2020ceb,Colleoni:2024knd,Hamilton:2021pkf,Thompson:2023ase}. Each whisker shows the 50\% and 90\% credible regions, with the circle marker being the median value. The horizontal dashed line in the last row represents the GR $A=1$ value, showing that all events are consistent with GR at the 90\% level. The surrogate posterior for GW150914 is from Ref.~\cite{Islam:2023zzj}.}
		\label{fig:MemoryPosteriors}
	\end{figure*}
}

\newcommand{\figTGR}{%
	\begin{figure}[t]
		\begin{center}
			\includegraphics[width=\columnwidth]{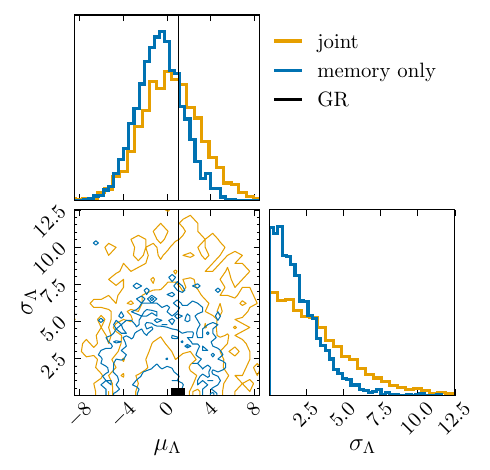}
		\end{center}
		\caption{Mean and standard deviation (see Eq.~\eqref{eq:hierarchicalLLfinal}) of the memory enhancement factor across the population of events up to the GWTC-5.0 catalog~\cite{LIGOScientific:2026wfs}. Contours represent the $1\sigma$, $2\sigma$, and $3\sigma$ credible regions. The yellow contours correspond to simultaneously fitting the memory hyperparameters and the astrophysical population parameters, while the blue contours correspond to fitting only the two memory hyperparameters. The black marker/line represents the GR prediction for the memory distribution: a delta function at 1---$\mu_{\Lambda}=1$, $\sigma_{\Lambda}=0$.}
		\label{fig:TGR}
	\end{figure}
}

\newcommand{\figForecast}{%
	\begin{figure*}[t]
		\begin{center}
			\includegraphics[width=\textwidth]{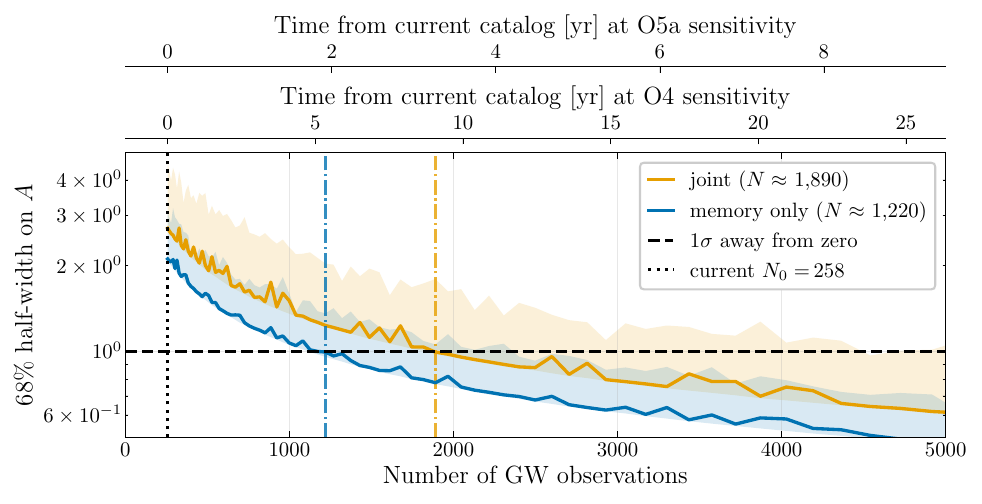}
		\end{center}
		\caption{Evolution of the 68\% half-width on $A$ as a function of the number of total GW observations. Yellow shows the projection for future O4-like observations using the 68\% half-widths on $\mu_{\Lambda}$ and $\sigma_{\Lambda}$ from the joint analysis on the current catalog of events, while blue shows the projection based on the 68\% half-widths from that of the memory-only analysis. The thick line for each of the two curves represents the median across many pseudo-catalog realizations, while the bands show the central 68\% spread. The vertical dotted line shows the current number of events analyzed: \Nmemoryevents. The horizontal dashed line shows the point at which the 68\% half-width on $A$ is constrained to be less than 1. The vertical dot-dashed lines show the number of observations required for the median 68\% half-width to cross 1, i.e., to constrain $A$ away from zero. The top axes show the observing time required to hit the number of GW observations shown based on the O4 and O5a rates of \OFourRates and \OFiveRates events per year~\cite{IGWN:ObservingCapabilities:2026}.
		}
		\label{fig:Forecast}
	\end{figure*}
}

\begin{document}

\author{Keefe Mitman
  \orcidlink{0000-0003-0276-3856}}
\email{kem343@cornell.edu} \Cornell
\author{Maximiliano Isi 
	\orcidlink{0000-0001-8830-8672}}
\email{msi2114@columbia.edu} \Columbia\CCA
\author{Will M. Farr
	\orcidlink{0000-0003-1540-8562}}
\email{will.farr@stonybrook.edu} \StonyBrook\CCA

\hypersetup{pdfauthor={Mitman et al.}}

\title{Constraining Gravitational Wave Memory with Hierarchical Inference}

\begin{abstract}
  With the multitude of gravitational wave observations that have been made in the past ten years, probing the dynamical and nonlinear nature of strong gravity is becoming more and more feasible. One promising way to test the nonlinear nature of Einstein's theory of general relativity (GR) is through the gravitational wave null memory effect: a nonlinear prediction of GR which corresponds to initially comoving observers being permanently displaced due to a burst of gravitational radiation. Previous studies have shown that, while it is unlikely that the memory effect will be observed in a single event by the LIGO-Virgo-KAGRA (LVK) detectors, evidence for memory in the population of LVK events should be attainable after $\sim$2,000 gravitational wave detections. These works, however, largely relied on Bayes factors to perform their memory analyses: an approach that can depend sensitively on the analysis priors and, when naively multiplied across many events, can even favor incorrect conclusions. In this work, using the GWTC-5.0 catalog of binary black hole observations, we instead perform hierarchical Bayesian inference---which is not subject to the issues associated with Bayes factors---to measure the evidence for memory in current LVK observations. We find that we can constrain what we call the memory enhancement factor---the constant appearing in front of the contribution to the strain from the supermomentum flux---to \Aconstraint (with $\pm$ values denoting the 68\% credible interval), consistent with its GR value of 1. We also forecast that $\sim$\NForecastCrossing detections will be needed to constrain the memory enhancement factor away from zero at the $1\sigma$ level.
\end{abstract}

\maketitle

\section{Introduction}
\label{sec:introduction}

A key motivation for gravitational wave (GW) science is to probe the rich phenomenology that is predicted by Einstein's theory of general relativity (GR) and to determine whether the gravitational wave interactions that we observe in nature are accurately described by it. While GR has been remarkably successful so far~\cite{Will:2014kxa,LIGOScientific:2026qni,LIGOScientific:2026fcf,LIGOScientific:2026wpt}, it is largely believed to be an incomplete theory, in part because it is not currently known how to reconcile it with quantum field theory in an ultraviolet-complete way~\cite{Goroff:1985th}. Consequently, an important objective for GW science is to scrutinize GR's consistency with nature by trying to measure unique predictions that have yet to be observed. In this regard, the gravitational wave null memory effect, because of its unique phenomenology and nonlinear origin, is an alluring unobserved effect to search for~\cite{Zeldovich:1974gvh,Christodoulou:1991cr,Blanchet:1992br,Wiseman:1991ss,PhysRevD.45.520}. 

The GW memory effect corresponds to the permanent net displacement that two initially comoving observers will experience due to a burst of gravitational radiation.\footnote{While there are also many other types of GW memory effects or, more generally, ``persistent observables'', in this work since we will exclusively focus on the \emph{displacement} memory, we will refer to this effect broadly as ``memory'' or the ``memory effect''~\cite{Cachazo:2014fwa,Pasterski:2015tva,Nichols:2018qac,Flanagan:2018yzh,Flanagan:2019ezo,Grant:2021hga,Grant:2023ged}.} Its so-called \emph{ordinary} form was originally found in Ref.~\cite{Zeldovich:1974gvh} in 1974 by studying hyperbolic black hole scattering, while its so-called \emph{null} form is commonly traced back to Ref.~\cite{Christodoulou:1991cr}, but was also noticed a few years earlier in Refs.~\cite{Payne:1983rrr,Blanchet1990HDR}.\footnote{Originally the two contributions to the memory were called ``linear'' and ``nonlinear'' (or ``Christodoulou''). However, they have since been renamed to ``ordinary'' and ``null'' to better reflect the way in which they are sourced~\cite{Bieri:2013ada}; see, e.g., Ref.~\cite{Mitman:2024uss} for a review.} The ordinary contribution to the memory is sourced by unbound sources and exists in linearized GR, while the null contribution is sourced by gravitational radiation interacting with itself and is thus inherently nonlinear.

 Apart from its impact on observers, the memory effect is also intriguing because it is intimately connected to asymptotic symmetries~\cite{Newman:1966ub,Ashtekar:1981bq,Geroch:1981ut,Dray:1984rfa,Ludvigsen:1989kg,Flanagan:2014kfa,Goncharov:2023woe} and soft theorems~\cite{Weinberg:1965nx,Strominger:2013jfa,Strominger:2014pwa}. As a result, memory plays an important role in fixing the coordinate freedom of gravitational systems~\cite{Mitman:2021xkq,Mitman:2022kwt}, formulating the concept of celestial holography~\cite{Strominger:2017zoo,Raclariu:2021zjz,Pasterski:2021raf,Pasterski:2021rjz}, and even in working toward establishing a divergence-free and self-consistent theory of quantum gravity~\cite{Prabhu:2022zcr}.

Unfortunately, because the memory effect is largely a low-frequency effect, it is nearly impossible to detect in individual binary black hole (BBH) observations with the current LIGO \cite{LIGOScientific:2014pky}, Virgo \cite{VIRGO:2014yos}, and KAGRA \cite{KAGRA:2020tym} (LVK) GW detectors~\cite{Lasky:2016knh}.  Nonetheless, Ref.~\cite{Lasky:2016knh} showed that one can measure the memory in a population of observations through a coherent summation of sub-threshold signals, often referred to as ``signal-to-noise ratio (SNR) stacking''. Assuming a population of GW150914-like events, which at the time was the only direct GW observation that had been made, Ref.~\cite{Lasky:2016knh} showed that 35 (90) events would be needed to achieve a combined memory SNR of 3 (5) with this method. Later, once a much more informative GW population had been established by the LVK detectors, Refs.~\cite{Hubner:2019sly,Boersma:2020gxx,Hubner:2021amk,Grant:2022bla,Cheung:2024zow}\footnote{Note that in an earlier version of the \textsc{gwmemory} package used by some of these works, a bug was present that led to the incorrect calculation of the memory~\cite{Talbot:2018sgr}. This has since been fixed.} conducted various studies on memory using LVK catalogs up to the GWTC-3.0 catalog and concluded that memory had yet to be measured in the population of events and would require $\mathcal{O}(2$,$000)$ events to do so. All of these analyses,\footnote{In addition to estimating Bayes factors, Ref.~\cite{Cheung:2024zow} also performed a hierarchical analysis on the memory contribution to the strain, parameterized in a similar manner to the analysis that we present in this work, using the GWTC-3.0 catalog. Their model required the memory be the same in every observation; this is the case in GR, but need not occur in beyond-GR theories. Our analysis allows one to learn the distribution of the memory across events, which is important for robustly detecting beyond-GR effects \cite{Isi:2019asy}.} however, relied on Bayes factors for this projection---an approach that can yield incorrect results when combining many events~\cite{Zimmerman:2019wzo,Isi:2019asy,Isi:2022cii}---and did not simultaneously infer the memory alongside the population of other astrophysical parameters, which is crucial for performing a consistent catalog analysis~\cite{Payne:2023kwj}.

In this work, we do exactly that. Namely, we utilize hierarchical Bayesian inference to simultaneously infer the evidence for the memory in the LVK Collaboration's GWTC-5.0 catalog of observations~\cite{LIGOScientific:2026wfs} with the usual astrophysical parameter population~\cite{LIGOScientific:2026ctl}. We structure the paper according to the following. In Sec.~\ref{sec:memory} we begin by reviewing how the memory can be computed from the GW strain and how this will be relevant to our analysis. In Sec.~\ref{sec:methodology} we illustrate our methodology for including memory in the analysis of the GWTC-5.0 catalog and then performing hierarchical inference to try to constrain memory across the population of GW events. In Sec.~\ref{sec:results} we provide our various results, with Sec.~\ref{sec:memoryconstraints} showing the constraints on the memory population obtained via hierarchical inference and Sec.~\ref{sec:forecasts} showing a forecast for how many GW events are required to measure memory. And, finally, in Sec.~\ref{sec:discussion} we summarize our findings and highlight prospects for observing memory in the future.

\section{Computation of memory}
\label{sec:memory}

Until recently, numerical relativity (NR) simulations of BBH mergers could not resolve the memory effect~\cite{Favata:2010zu,Pollney:2010hs,Mitman:2020pbt}. This was largely because the way the strain waveforms were extracted to future null infinity in NR, a technique known as extrapolation~\cite{Boyle:2019kee}, was insufficient for resolving the low-frequency memory effect.\footnote{The more robust method of using Cauchy-characteristic evolution (CCE) for obtaining waveforms at future null infinity, however, does resolve memory effects~\cite{Bishop:1996gt,Bishop:1997ik,Bishop:1998uk,Reisswig:2009rx,Reisswig:2009us,Babiuc:2010ze,Pollney:2010hs,Handmer:2014qha,Handmer:2015dsa,Handmer:2016mls,Moxon:2020gha,Moxon:2021gbv,Mitman:2020pbt}.} As a result of this--- at least in part---the majority of strain waveform models do not contain memory~\cite{Taracchini:2013rva,Pan:2013rra,Hannam:2013oca,Bohe:2016gbl,Ossokine:2020kjp,Varma:2019csw,Pratten:2020ceb,Colleoni:2024knd,Pompili:2023tna,Ramos-Buades:2023ehm,Hamilton:2021pkf,Thompson:2023ase}.\footnote{See Ref.~\cite{Yoo:2023spi} for a model trained on CCE waveforms which does contain memory effects.} Fortunately, due to the way in which the memory is sourced, for any model of the GW strain $h=h_{+}-ih_{\times}$, the memory can be computed from that model in a straightforward manner~\cite{Talbot:2018sgr,Mitman:2020pbt,Mitman:2020bjf}.
\begin{widetext}
\noindent In particular, motivated by the Bondi-van der Burg-Metzner-Sachs (BMS) symmetry group and its corresponding charges and fluxes~\cite{Wald:1999wa,Flanagan:2015pxa,Mitman:2020pbt}, one can write the real part of the GW strain as
\begin{align}
	\label{eq:restrain}
	\mathrm{Re}\left[h\right]=\bar{\eth}^{2}\mathfrak{D}^{-1}\left[m+\frac{1}{4}\int_{-\infty}^{u}|\dot{h}|^{2}\,{\rm d}u\right],
\end{align}
for $\eth$ the spin-weight operator~\cite{Newman:1961qr,Goldberg:1966uu,Geroch:1973am} whose action on spin-weighted spherical harmonics~\cite{Goldberg:1966uu} is
\begin{align}
	\label{eq:eth}
	\eth\phantom{}_{s}Y_{\ell,m}=\frac{1}{\sqrt{2}}\sqrt{(\ell-s)(\ell+s+1)}\phantom{}_{s+1}Y_{\ell,m},
\end{align}
$\mathfrak{D}$ an operator whose action on spherical harmonics is
\begin{align}
	\label{eq:Dfrak}
	\mathfrak{D}Y_{\ell,m}=\frac{1}{8}(\ell+2)(\ell+1)\ell(\ell-1)Y_{\ell,m},
\end{align}
$m$ the Bondi mass aspect~\cite{Bondi:1962px}, and overdots representing derivatives with respect to the Bondi time $u\equiv t-r$. Thus, since the null memory is sourced by the supermomentum flux---the second term in Eq.~\eqref{eq:restrain}---one can compute the null memory expressed in the gravitational wave strain as the $u\rightarrow+\infty$ limit\footnote{Formally, one must also have that the spacetime is non-radiative as $u\rightarrow\pm\infty$ for the memory to be well defined.} of
\begin{align}
	\label{eq:memory}
	h_{\mathrm{mem}}=\frac{1}{4}\bar{\eth}^{2}\mathfrak{D}^{-1}\int_{-\infty}^{u}|\dot{h}|^{2}\,{\rm d}u.
\end{align}
Critically, the computation of $h_{\mathrm{mem}}$ requires knowledge of $h$ (or really its time derivative $\dot{h}$) over the two-sphere~\cite{Wiseman:1991ss,PhysRevD.45.520}. This is because the differential operator in front of the integral in Eq.~\eqref{eq:memory} is not local in the $(\theta,\phi)$ angles on the two-sphere, as highlighted by Eqs.~\eqref{eq:eth} and~\eqref{eq:Dfrak}. Consequently, the memory cannot be constructed from the waveform at a single point on the sky; instead one must provide the strain waveform in terms of an angular basis, e.g., the spin-weight $s=-2$ spherical harmonics. To express the memory contribution $h_{\mathrm{mem}}$ from Eq.~\eqref{eq:memory} in terms of spin-weight $s=-2$ spherical harmonics, one can use the integral formula
\begin{align}
	\int_{S^{2}}\phantom{}_{s_{1}}Y_{\ell_{1},m_{1}}\phantom{}_{s_{2}}Y_{\ell_{2},m_{2}}\phantom{}_{s_{3}}Y_{\ell_{3},m_{3}}\,{\rm d}\Omega=\sqrt{\dfrac{(2\ell_{1}+1)(2\ell_{2}+1)(2\ell_{3}+1)}{4\pi}}\begin{pmatrix}\ell_{1}&\ell_{2}&\ell_{3}\\m_{1}&m_{2}&m_{3}\end{pmatrix}\begin{pmatrix}\ell_{1}&\ell_{2}&\ell_{3}\\-s_{1}&-s_{2}&-s_{3}\end{pmatrix},
\end{align}
for ${\rm d}\Omega$ the volume form on the two-sphere and the rightmost objects the Wigner-3j symbols with $s_{1}+s_{2}+s_{3}=0$. Equipped with this, one can compute the memory contribution to the individual $(\ell,m)$ spin-weight $s =-2$ modes as
\begin{align}
	\label{eq:memorymodes}
	h_{\mathrm{mem}}^{\ell,m}(u)&=\sum\limits_{\substack{\ell_{1},\ell_{2}\\m_{2}}}(-1)^{m_{2}}\sqrt{\dfrac{(2\ell_{1}+1)(2\ell_{2}+1)(2\ell+1)}{4\pi(\ell+2)(\ell+1)\ell(\ell-1)}}\begin{pmatrix}\ell_{1}&\ell_{2}&\ell\\m-m_{2}&m_{2}&-m\end{pmatrix}\begin{pmatrix}\ell_{1}&\ell_{2}&\ell\\-2&2&0\end{pmatrix}\int_{-\infty}^{u}\dot{h}^{*}_{\ell_{1},m-m_{2}}\dot{h}_{\ell_{2},m_{2}}\,{\rm d}u,
\end{align}
\end{widetext}
where $\phantom{}^{*}$ on $\dot{h}_{\ell_{1}, m-m_{2}}$ denotes complex conjugation. Thus, for a waveform model of the strain which provides the spin-weight $s =-2$ modes, one can readily compute the memory contribution to each mode using Eq.~\eqref{eq:memorymodes}.

Importantly, Ref.~\cite{Mitman:2020bjf} showed that the NR waveforms produced using extrapolation which do not exhibit the memory effect can be corrected to much better match the NR waveforms which do exhibit the memory effect by simply adding the time-dependent $h_{\mathrm{mem}}$ from Eq.~\eqref{eq:memory}, thereby making these memoryless NR waveforms more consistent with GR.\footnote{Another way to view this ``memory correction'' of Ref.~\cite{Mitman:2020bjf} is as a post-processing procedure that makes the NR waveforms better satisfy the $\Psi_{2}$ Bianchi identity, which is just the time derivative of Eq.~\eqref{eq:restrain}. While this clearly does not capture all of the physics that may be missing, such as the memory of the memory (which we stress is a very small contribution), it is certainly an important first-order improvement to memoryless waveforms.} Thus, since waveform models mimic these extrapolated waveforms, we can correct these models to better match the predictions of GR by adding Eq.~\eqref{eq:memory} to their predictions. With these GR-consistent waveforms, we can then proceed to try to constrain the memory.

An important caveat regarding observing the memory, however, is that formally the memory is really only defined as a difference in the strain between non-radiative regimes of future null infinity, e.g., between the two boundaries $u\rightarrow-\infty$ and $u\rightarrow+\infty$. Ideally, to observe the memory we would design an analysis which is only sensitive to a net change in the strain, e.g., measuring the net change in the time-averages of the strain in some window before and after the merger phase~\cite{Scargle:2021lqz}.\footnote{While this paper has since been withdrawn, the analysis that was presented in this work is similar to what we mention here.} However, since this type of analysis is largely agnostic to the underlying GW signal, it will tend to provide weak constraints. Alternatively, one can instead choose a \emph{time evolution} for the memory and try to constrain the presence of this contribution in the observed GW signal. But, how one should represent the memory's time evolution is not obvious. Put differently, there is no meaningful way to define the instantaneous memory at some finite time $u$. For example, one could take the time evolution of the memory 
\begin{enumerate}
	\item to be a step function, which is a strong assumption since it corresponds to a discontinuous signal (and, as a consequence, can likely already be ruled out by current observations as not being present);
	\item to be that of Eq.~\eqref{eq:memory}, which although agrees with the memory when taking its difference between the boundaries $u\rightarrow-\infty$ and $u\rightarrow+\infty$ may include high-frequency content in the intermediate regime;
	\item or, to be some kind of Isaacson average of Eq.~\eqref{eq:memory}, which assumes the memory's time evolution should be monotonic, which is also not obvious~\cite{Zosso:2026czc}.
\end{enumerate}
With one of these choices, one can try to measure this proxy for the memory by studying the strain model
\begin{align}
	\label{eq:strainmodel}
	h(u;\,\lambda,A)=h_{\mathrm{no\,mem}}(u;\lambda) + A\,h_{\mathrm{mem}}(u;\lambda) \, ,
\end{align}
for $\lambda$ the 15 intrinsic and extrinsic parameters\footnote{Note that the binary black hole parameter space is really described by a 17-dimensional vector. However, since our waveform models do not currently include the eccentricity and mean anomaly, the $\lambda$ that we use is only 15-dimensional.} and $A$ the ``memory enhancement factor'', which is 1 in GR. Equation~\eqref{eq:strainmodel} can then be used to measure $A$ across the population of GW observations, by taking $h_{\mathrm{no\,mem}}$ to be the GR-consistent waveform minus $h_{\mathrm{mem}}$ and $h_{\mathrm{mem}}$ the proxy for the time-dependent evolution of the memory. While there are various arguments for the last two proxy choices presented earlier, in this analysis we choose to use the proxy described by point (ii), as it is well-motivated from the BMS charge and flux decomposition and because $h_{\mathrm{no\,mem}}$ for (ii) already corresponds to the waveforms used in the LVK analysis. Therefore, in the remainder of this work we will use Eq.~\eqref{eq:strainmodel} to try to measure the memory, with $h_{\mathrm{no\,mem}}$ the waveform used in the LVK analysis, i.e., that of Refs.~\cite{Varma:2019csw,Pratten:2020ceb,Colleoni:2024knd,Pompili:2023tna,Ramos-Buades:2023ehm,Hamilton:2021pkf,Thompson:2023ase}, and $h_{\mathrm{mem}}$ the expression in Eq.~\eqref{eq:memory}, or the sum of modes from Eq.~\eqref{eq:memorymodes}, using $h=h_{\mathrm{no\,mem}}$.

Last, the memory analysis based on Eq.~\eqref{eq:strainmodel} which we will subsequently perform is not expected to represent the structure of any well-formulated beyond-GR theory, since we do not expect there to be any beyond-GR theory which preserves $h_{\mathrm{no\,mem}}$ and $h_{\mathrm{mem}}$ while uniquely predicting some $A\not=1$. While certain beyond-GR theories predict the memory contribution to the strain to be different from that predicted by GR~\cite{Lang:2013fna,Du:2016hww,Koyama:2020vfc,Tahura:2021hbk,Bernard:2022noq,Heisenberg:2023prj,Tahura:2025ebb,Heisenberg:2025roe,Heisenberg:2025tfh,Gasparotto:2026bru,Zosso:2026uty}, these theories also modify the non-memory and memory parts of the strain. That being said, an analysis using Eq.~\eqref{eq:strainmodel} on waveforms predicted by a beyond-GR theory would likely result in a measurement of $A\not=1$, but this should instead be interpreted as a bias in the inference, in the same way that the black hole progenitor parameters would be biased due to the waveform model not matching the theory that produced the observed GW. Put differently, this analysis can be used as a straightforward null test of GR.
\section{Methodology}
\label{sec:methodology}

\subsection{Memory computation}

As stated in Sec.~\ref{sec:memory}, standard LVK analyses do not use waveform models that contain the expected contribution from Eq.~\eqref{eq:memory}, which can be used as a proxy for memory. For such an analysis, the log-likelihood used to produce the public posteriors of parameters $\lambda$ takes the form
\begin{align}
	\log\mathcal{L}&=-\frac{1}{2}\left\langle d- R(\lambda)\, h(\lambda) \mid d-R(\lambda)\, h(\lambda)\right\rangle+C\nonumber\\
	&=-\frac{1}{2}\left\langle d-\tilde{h}(\lambda) \mid d-\tilde{h}(\lambda)\right\rangle+C\nonumber\\
	&=-\frac{1}{2}\left\|d-\tilde{h}(\lambda)\right\|^2+C,
\end{align}
where $d$ is the observed data, $h(\lambda)$ is the waveform model evaluated at parameters $\lambda$, $R(\lambda)$ is the detector response which includes the calibration and antenna patterns, $\tilde{h}(\lambda)\equiv R(\lambda)h(\lambda)$ is the signal at the detector,
\begin{align}
	\langle h_{1} | h_{2} \rangle \equiv 4 \mathrm{Re}\left[\int_0^\infty h_{1}^{*}(f) h_{2}(f)/S(f)\,{\rm d}f\right]
\end{align}
is the frequency-domain noise-weighted inner product with some noise power spectral density $S(f)$, $\left\| \cdot \right\|^2\equiv\langle \cdot | \cdot \rangle$, and $C$ is a constant independent of the $\lambda$ parameters. Letting $r(\lambda)\equiv d-\tilde{h}(\lambda)$ denote the residual between the data and the template, the log-likelihood simplifies to
\begin{align}
	\log\mathcal{L}=-\frac{1}{2}\left\|r(\lambda)\right\|^2+C.
\end{align}

If we wish to include the missing contribution to the waveform models and thus also the effects of memory, since the memory can be computed directly from the usual waveform models using Eqs.~\eqref{eq:memorymodes} and~\eqref{eq:memory}, the log-likelihood becomes
\begin{align}
	\label{eq:memoryLLpre}
	\log\mathcal{L}=-\frac{1}{2}\left\| r(\lambda)-\tilde{h}_{\mathrm{mem}}(\lambda)\right\|^2+C,
\end{align}
where $\tilde{h}_{\mathrm{mem}}$ is $R(\lambda)h_{\mathrm{mem}}(\lambda)$ for $R(\lambda)$ the same $R(\lambda)$ as was used in the original analysis. Crucially, this requires no new waveform model. But, if we wish to instead \emph{detect} the memory, then in Eq.~\eqref{eq:memoryLLpre} we want to include the memory enhancement factor via
\begin{align}
	\label{eq:memoryLL}
	\log\mathcal{L}=-\frac{1}{2}\left\| r(\lambda)-A\,\tilde{h}_{\mathrm{mem}}(\lambda)\right\|^2+C.
\end{align}
Computing the posterior on $A$ allows us to easily test the consistency of the observed memory effect with the predictions of GR, and to determine whether the data prefer a non-zero memory effect at all.
Rather than sampling this parameter directly, which would require reanalyzing the data from scratch, we can take advantage of the structure of Eq.~\eqref{eq:memoryLL} to do this efficiently.

Conveniently, Eq.~\eqref{eq:memoryLL} takes the form of a Gaussian on $A$, which peaks at the value
\begin{align}
	\label{eq:mean}
	\mu=\dfrac{\mathrm{Re}\left[\langle\tilde{h}_{\mathrm{mem}}(\lambda)|r(\lambda)\rangle\right]}{\left\|\tilde{h}_{\mathrm{mem}}(\lambda)\right\|^2},
\end{align}
with uncertainty
\begin{align}
	\label{eq:stddev}
	\sigma=\left\|\tilde{h}_{\mathrm{mem}}(\lambda)\right\|^{-1}.
\end{align}
Marginalizing out the memory enhancement factor of the likelihood in Eq.~\eqref{eq:memoryLL} with a flat prior therefore yields the marginal log-likelihood
\begin{align}
	\label{eq:marginalLL}
	\log\overline{\mathcal{L}}&=\log\int\exp(\log\mathcal{L})\,{\rm d} A\nonumber\\&=-\frac{1}{2}\left(||r(\lambda)||^{2}-\sigma^{-2}\mu^{2}\right)-\frac{1}{2}\log\left(2\pi\sigma^{-2}\right)+C.
\end{align}
Thus, starting from posterior samples from the standard memoryless LVK analysis, we can derive a conditional and marginal posterior for the memory parameter $A$.

The sample-level hyperparameters $\mu$ and $\sigma$ respectively yield the preferred $A$ value as well as our uncertainty about this value, conditioned on the corresponding $\lambda$ sample. Meanwhile, the marginal likelihood $\overline{\mathcal{L}}$ in Eq.~\eqref{eq:marginalLL} encodes the goodness of fit achieved by incorporating memory in the analysis. Therefore, if we have a set of samples drawn from a posterior based on the original, no-memory likelihood $\mathcal{L}$, then the ratio of the marginal likelihood to the original likelihood serves as an importance weight to resample draws from the memoryless posterior into the posterior that would have been obtained with memory.
The log-weight is simply
\begin{align}
	\label{eq:weight}
	\log w&=\log\overline{\mathcal{L}}(\lambda)-\log\mathcal{L}(\lambda)\nonumber\\&=\frac{1}{2}\sigma^{-2}\mu^{2}-\frac{1}{2}\log\left(2\pi\sigma^{-2}\right).
\end{align}
If the memory is not very strong in any individual event, then the importance weights will be very close to 1, and the re-weighted samples will remain close to the originals. Furthermore, each sample can also be augmented by a value of $A$ that is drawn from the Gaussian distribution described by Eqs.~\eqref{eq:mean} and~\eqref{eq:stddev} to obtain a new set of samples that were instead drawn from the posterior on $A$. This procedure is similar to the likelihood-reweightings that were performed previously in, e.g., Refs.~\cite{Romero-Shaw:2019itr,Payne:2019wmy}.

With the values of $\mu$ and $\sigma$ obtained for each sample, we can then combine them in a hierarchical way to obtain an overall posterior on the population distribution of $A$ across the entire catalog of GW events to determine whether the data prefer a non-zero memory effect in the population as a whole~\cite{Isi:2022cii}.

\subsection{Hierarchical inference}
\label{sec:hierarchicalinference}

Following the work of Ref.~\cite{Payne:2023kwj}, we simultaneously model the population distribution of the memory and all of the usual BBH merger astrophysical parameters. We assume that we have regular parameter estimation samples $\lambda_{s}$ that have been enhanced to include both $\mu$ and $\sigma$ from Eqs.~\eqref{eq:mean} and~\eqref{eq:stddev}, which we will hereafter call $\mu_{s}$ and $\sigma_{s}$ when referring to the sample $s$. For each of our samples, this defines the conditional distribution on the memory enhancement factor $A$ given $\lambda_{s}$, which is
\begin{align}
	\label{eq:Aconditionaldistribution}
	p(A|\lambda_{s})=\mathcal{N}(A \mid \mu_{s},\sigma_{s}),
\end{align}
where $\mathcal{N}(A \mid \mu_{s}, \sigma_{s})$ is a normal probability density that is evaluated at $A$.

The hierarchical likelihood for population parameters $\Lambda$ for the $i^{\mathrm{th}}$ event requires computing the integral
\begin{align}
	\label{eq:hierarchicalLL}
	\mathcal{L}_{i}=\int\dfrac{p(A,\lambda)}{W(\lambda)}\, p(A,\lambda \mid \Lambda)\,{\rm d} A {\rm d}\lambda,
\end{align}
where $p(A,\lambda)$ is the sample posterior and $W(\lambda)$ is the importance weight that encodes the sampling prior from the original analysis and any of the necessary Jacobians, as well as the inverse of the importance weight that was derived in Eq.~\eqref{eq:weight}; we will assume that the prior on $A$ is flat so the weight $W$ in Eq.~\eqref{eq:hierarchicalLL} is independent of $A$. Because $p(A,\lambda)=p(A | \lambda)\, p(\lambda)$, Eq.~\eqref{eq:hierarchicalLL} becomes
\begin{align}
	\label{eq:hierarchicalLLpartial}
	\mathcal{L}_{i}=\int \dfrac{p(A \mid \lambda)\, p(\lambda)}{W(\lambda)}\, p(A,\lambda\mid\Lambda)\,{\rm d}A {\rm d}\lambda.
\end{align}
At this point, we can use the fact that we have $N$ samples from the posterior $\lambda_{s}$, i.e.,
\begin{align}
	p(\lambda)\approx\frac{1}{N}\sum\limits_{s}\delta(\lambda-\lambda_{s})
\end{align}
to approximate Eq.~\eqref{eq:hierarchicalLLpartial} with a Monte-Carlo average as
\begin{align}
	\mathcal{L}_{i}&\approx\frac{1}{N}\sum_{s}\int\dfrac{p(A \mid \lambda)}{W(\lambda)}\, p(A,\lambda \mid \Lambda)\, \delta(\lambda-\lambda_{s})\,{\rm d}A {\rm d}\lambda\nonumber\\
	&=\frac{1}{N}\sum_{s}\int\dfrac{p(A \mid \lambda_{s})}{W(\lambda_{s})}\, p(A,\lambda_{s} \mid \Lambda)\, {\rm d}A.
\end{align}
If we further assume that the population model factorizes as $p(A,\lambda_{s} \mid \Lambda)=p(A \mid \Lambda_{A})\, p(\lambda_{s} \mid \Lambda_{\lambda})$ where $\Lambda_{A}$ and $\Lambda_{\lambda}$ are the memory and astrophysical hyperparameters, then
\begin{align}
	\label{eq:hierarchicalLLsimplified}
	\mathcal{L}_{i}\approx\frac{1}{N}\sum\limits_{s}\dfrac{p(\lambda_{s} \mid \Lambda_{\lambda})}{W(\lambda_{s})}\int p(A \mid \lambda_{s})\, p(A \mid \Lambda_{A})\,{\rm d}A.
\end{align}
To further simplify this expression, we need to assume a specific population model for the memory parameter $A$.

To establish whether there is evidence for memory in the GW population as predicted by GR, we want to ascertain whether $A = 1$ identically for all systems, i.e., whether the population distribution for $A$ is simply $p(A \mid \Lambda_A) = \delta(A - 1)$. In alternative scenarios, we might expect that there is no memory, i.e., $p(A \mid \Lambda) = \delta(A)$, or that different systems manifest different degrees of memory relative to the default GR prediction, resulting in some arbitrary distribution of $A$ with some variance.
To capture the mean and variance of the distribution of $A$, we follow Ref.~\cite{Isi:2019asy} and set our population model for $A$ to be a Gaussian of arbitrary mean $\mu_\Lambda$ and scale $\sigma_\Lambda$, such that $p(A \mid \Lambda_A) = \mathcal{N}(A \mid \mu_\Lambda, \sigma_\Lambda)$ and $\Lambda_A = \{ \mu_\Lambda, \, \sigma_\Lambda\}$.
This model serves as a null test, by which consistency with $(\mu_\Lambda, \sigma_\Lambda)=(1, 0)$ is required for agreement with the GR expectation of $A=1$.

\figMemoryPosteriors

Under this Gaussian population model for the memory, the integral over the memory parameter $A$ in Eq.~\eqref{eq:hierarchicalLLsimplified} can be performed analytically, yielding
\begin{align}
	\label{eq:hierarchicalLLfinal}
	\mathcal{L}_{i}&\approx\frac{1}{N}\sum\limits_{s}\dfrac{p(\lambda_{s} \mid \Lambda_{\lambda})}{W(\lambda_{s})} \hspace{-3pt} \int\mathcal{N}(A \mid \mu_{s},\sigma_{s})\, \mathcal{N}(A \mid \mu_{\Lambda},\sigma_{\Lambda})\,{\rm d}A\nonumber\\
	&=\frac{1}{N}\sum\limits_{s}\dfrac{p(\lambda_{s} \mid \Lambda_{\lambda})}{W(\lambda_{s})}\, \mathcal{N}\left(\mu_{s}\,\Big|\, \mu_{\Lambda},\sqrt{\sigma_{s}^{2}+\sigma_{\Lambda}^{2}}\right),
\end{align}
where we used Eq.~\eqref{eq:Aconditionaldistribution} to replace $p(A \mid \lambda_{s})$ in Eq.~\eqref{eq:hierarchicalLLsimplified}. Equation~\eqref{eq:hierarchicalLLfinal} thus provides a straightforward way to evaluate the hierarchical likelihood \emph{without sampling the latent variable $A$} for each sample. Instead, each sample contributes to the likelihood through a Gaussian term that compares its sample-level memory mean $\mu_{s}$, Eq.~\eqref{eq:mean}, with the memory population mean $\mu_{\Lambda}$, with a variance that is a sum in quadrature of the sample-level memory measurement uncertainty $\sigma_{s}$, Eq.~\eqref{eq:stddev}, and the overall memory population uncertainty $\sigma_{\Lambda}$.

Finally, while Eq.~\eqref{eq:hierarchicalLLfinal} provides the likelihood for an individual event, the log-likelihood for a catalog of some number of $N$ events is, up to a constant, given by 
\begin{align}
	\label{eq:poplike}
	\log\mathcal{L}(\Lambda) = \sum_{i=1}^{N}\log\mathcal{L}_{i}(\Lambda) - N \log \xi(\Lambda),
\end{align}
where $\xi(\Lambda)$ is the detection efficiency, which encodes the detection biases \cite{Mandel:2018mve}.
We assume there is no selection based on the memory parameter $A$, given that we expect the memory piece to have a particularly small effect on the signal-to-noise ratio \cite{Magee:2023muf}.
The hyperposterior on both $\mu_{\Lambda}$ and $\sigma_{\Lambda}$, obtained by sampling Eq.~\eqref{eq:poplike} together with some hyperprior $p(\mu_{\Lambda}, \sigma_{\Lambda})$, encodes our best knowledge of the most likely value of $A$ and the degree of variance consistent with the observed population.

For the astrophysical part of the model, i.e., $p(\lambda \mid \Lambda_\lambda)$, we adopt a parametric form based on the LVK analysis of GWTC-5.0 that was performed in Ref.~\cite{LIGOScientific:2026ctl}. We factorize the BBH parameter density for the primary mass $m_1$, mass ratio $q = m_2/m_1$, redshift $z$, spin magnitudes $\chi_{1/2}$ and spin tilts $\theta_{1/2}$. The primary mass $m_{1} \in [3, 300]\,M_{\odot}$ is described by a three-component mixture of a continuous broken power law and two Gaussian peaks truncated to the same support, with mixture weights drawn from a flat Dirichlet prior. The mass ratio follows the conditional power law $p(q \mid m_{1}) \propto q^{\beta}$ on $[3\,M_{\odot}/m_{1},\,1]$, while the redshift distribution scales as $(1+z)^{\lambda_{z}}\,\mathrm{d}V_{c}/\mathrm{d}z\,(1+z)^{-1}$ for $V_{c}$ the comoving volume in a Planck 2015 cosmology~\cite{Planck:2015fie}. The cosine spin tilts $(\cos\theta_{1}, \cos\theta_{2})$ are jointly modeled as a mixture of an isotropic component and a product of truncated Gaussians on $[-1, 1]$ with shared means and widths, which follows the non-independently but identically-distributed prescription used for GWTC-5.0; each spin magnitude $\chi_{i}$ is independently drawn from a Gaussian truncated to $[0, 1]$ with shared mean and width. We place broad uniform priors on these hyperparameters and account for potential selection effects through the rate-marginalized factor $\xi(\Lambda)^{-N}$ (see Eq.~\eqref{eq:poplike}), which is estimated as a Monte-Carlo average over the cumulative LVK simulated-injection campaign for GWTC-5.0~\cite{Essick:2025zed,LVK:GWTC5CumulativeSensitivity}. See Appendix \ref{sec:appendix-popmodel} for more details.

We draw samples from the hyperposterior by using the No-U-Turn Hamiltonian Monte Carlo algorithm~\cite{Hoffman:2014NUTS} as implemented in \textsc{NumPyro}~\cite{bingham2019pyro,Phan:2019NumPyro}.

\section{Results}
\label{sec:results}

For our memory analysis, we consider all BBH events observed by the LVK Collaboration that had an inverse false alarm rate above one year, up to and including the GWTC-5.0 catalog~\cite{LIGOScientific:2026wfs,LIGOScientific:2026ctl}. This amounted to \Nmemoryevents events in total.\footnote{Following the GWTC-5.0 population analysis, we exclude ER16 candidates from the hierarchical population inference~\cite{LIGOScientific:2026ctl}.} For each event, each available waveform model, and its corresponding posterior samples, we computed the memory mean $\mu_{s}$ and standard deviation $\sigma_{s}$ using Eqs.~\eqref{eq:mean} and~\eqref{eq:stddev}. The waveform models used are those described in Refs.~\cite{Varma:2019csw,Pratten:2020ceb,Colleoni:2024knd,Pompili:2023tna,Ramos-Buades:2023ehm,Hamilton:2021pkf,Thompson:2023ase}. Because the computation of the memory requires the strain modes per Eq.~\eqref{eq:memorymodes}, each waveform model was always used to provide as many modes as available. For the phenomenological models, which output waveforms in the frequency domain, we first transformed the output waveform to the time domain before computing the memory.

In Fig.~\ref{fig:MemoryPosteriors} we show the distribution of $\mu_{s}$ and $\sigma_{s}$ across all posterior samples for six of the more informative and popular BBH detections. As shown (especially by the third row of the figure), of these more significant events, none of them yield a confident measurement of memory and all have a preferred memory enhancement factor consistent with the GR value of $A = 1$, at least at the 90\% credible level. Across all of the analyzed events, GW250114~\cite{LIGOScientific:2025rid,LIGOScientific:2025wao} has the largest influence on the inference of the memory in the sense that when we remove this event from the hierarchical inference the posterior on $(\mu_{\Lambda},\sigma_{\Lambda})$ changes the most. As we will show in the subsequent section, even though no single event yields a confident measurement of memory, by instead inferring the memory hyperparameters across all events, we can obtain much better memory constraints.

\subsection{Memory Constraints}
\label{sec:memoryconstraints}

\figTGR

\figForecast

With the computation of $\mu_{s}$ and $\sigma_{s}$ performed for the posterior samples of all events and all of the available waveform models, we performed two hierarchical analyses: one in which we measured the memory hyperparameters $\mu_{\Lambda}$ and $\sigma_{\Lambda}$ (see Eq.~\eqref{eq:hierarchicalLLfinal}) in addition to all of the other astrophysical parameters, and one in which we measured only $\mu_{\Lambda}$ and $\sigma_{\Lambda}$, leaving the astrophysical population model fixed to the priors that are used in the single-event parameter estimation analyses. When conducting these analyses, we opted to use the following hierarchy for which waveform model to use. If available, we ideally used the NR surrogate, then the effective-one-body (EOB) models, and then the phenomenological (or Phenom) models.\footnote{For certain samples, we found that the waveforms produced by the Phenom models had unphysical spikes at low frequencies that caused anomalously large $\mu_{s}$ values. Because of this, when using the Phenom models, if $|\mu_{s}/\sigma_{s}|$ was $>10$ for any sample, we simply removed that sample from the analysis. This amounts to removing only a few tens of samples from the whole analysis.} This choice was made based on the model accuracy results shown in Fig.~2 of Ref.~\cite{LIGOScientific:2025rsn}, which yielded this ranking. When we used an EOB model, we used \texttt{SEOBNRv5PHM} if available, and otherwise \texttt{SEOBNRv4PHM}~\cite{Bohe:2016gbl,Ossokine:2020kjp,Pompili:2023tna,Ramos-Buades:2023ehm}. When we used a Phenom model, we used \texttt{IMRPhenomXO4a} if available, and otherwise \texttt{IMRPhenomPHM\_SpinTaylor}, or \texttt{IMRPhenomPHM}~\cite{Pratten:2020ceb,Colleoni:2024knd,Hamilton:2021pkf,Thompson:2023ase}.

Our results for this study are highlighted in Fig.~\ref{fig:TGR}, which shows the $1\sigma$, $2\sigma$, and $3\sigma$ credible regions for the posterior of $\mu_{\Lambda}$ and $\sigma_{\Lambda}$. As can be seen, across all of the available events that we analyzed, we can currently constrain the memory $\mu_{\Lambda}$ and $\sigma_{\Lambda}$ hyperparameters to \muLambdaconstraint and \sigmaLambdaconstraint. These values represent the mean and the 68\% credible interval of each of these variables. They are both obtained from the joint inference analysis. With these, we can compute our main constraint on $A$ by drawing one sample from $\mathcal{N}(\mu_{\Lambda},\sigma_{\Lambda})$ for each of the hypersamples of $(\mu_{\Lambda},\sigma_{\Lambda})$. This yields an $A$ of \Aconstraint. Therefore, we find that, as was expected, with the current catalog of GW observations we cannot claim a detection of memory as $A$ is not strongly bounded away from 0.\footnote{While performing this analysis, we noticed that a few events have a preference for large $A$ values. We find that this is due to using too small of an apodizing window, which leads to spectral leakage exciting low frequencies. By using a longer window, we find that this problem no longer persists. This likely also explains the large $A$ values that Ref.~\cite{Cheung:2024zow} similarly identified, but attributed to unmodeled non-Gaussian noise in the detectors.}

\subsection{Forecasts}
\label{sec:forecasts}

To forecast the number of events that will be needed to constrain $A$ away from zero, i.e., to confidently measure the memory, we will construct mock catalogs of future GW events by performing the following analysis. First, we note that from the hierarchical inference performed in Sec.~\ref{sec:memoryconstraints}, we have a set of hypersamples $\{\mu_{\Lambda}^{(k)},\sigma_{\Lambda}^{(k)}\}$ that were inferred from the current $N_{0}=\text{\Nmemoryevents}$ observations. And, for a future catalog of size $N$, the uncertainty on both $\mu_{\Lambda}$ and $\sigma_{\Lambda}$ will shrink as $1/\sqrt{N}$. Therefore, for some pseudo-catalog realization indexed by $r$, we can draw a catalog-level estimate for our hyperparameters around their GR values as
\begin{align}
	\hat{\mu}_{r}\sim\mathcal{N}\left(1,s_{\mu}\right)\quad\text{and}\quad
	\hat{\sigma}_{r}^{2}\sim\mathcal{N}\left(0,s_{\sigma^{2}}\right),
\end{align}
where
\begin{subequations}
\begin{align}
	s_{\mu}&\equiv\Delta \mu_{0}\sqrt{\frac{N_{0}}{N}},\\
	s_{\sigma^{2}}&\equiv\Delta \sigma_{0}^{2}\frac{N_{0}}{N}
\end{align}
\end{subequations}
for $\Delta\mu_{0}$ and $\Delta\sigma_{0}^{2}$ the 68\% half-widths computed from the current $\mu_{\Lambda}$ and $\sigma_{\Lambda}^{2}$ posteriors. From these, we can then draw hyperposterior samples for this pseudo-catalog via
\begin{subequations}
\begin{align}
	\mu_{\Lambda,r}&\sim\mathcal{N}\left(\hat{\mu}_{r},s_{\mu}\right),\\
	\sigma_{\Lambda,r}^{2}&\sim\mathrm{InvGamma}\left(\alpha,\beta\right)
\end{align}
\end{subequations}
for
\begin{subequations}
\begin{align}
	\alpha&\equiv2+\frac{\hat{\sigma}_{r}^{4}}{s_{\sigma^{2}}^{2}},\\
	\beta&\equiv\hat{\sigma}_{r}^{2}(\alpha-1).
\end{align}
\end{subequations}
The inverse-gamma distribution used to represent $\sigma_{\Lambda,r}^{2}$ is constructed in this way so that it corresponds to a scaled inverse-$\chi^{2}$ distribution whose expectation value and standard deviation match $\hat{\sigma}_{r}^{2}$ and $s_{\sigma^{2}}$. With each $\mu_{\Lambda,r}$ and $\sigma_{\Lambda,r}^{2}$ sample, we now draw a value for $A_{r}$, i.e., the pseudo-catalog realization of $A$, as $A_{r}\sim\mathcal{N}\left(\mu_{\Lambda,r},\sigma_{\Lambda,r}\right)$. And, finally, across all of the $A_{r}$ draws, we can measure a 68\% half-width $\Delta A_{r}$ to understand how our uncertainty on $A$ changes as we observe more and more GW events. By repeating this over many pseudo-catalog realizations, we can obtain an ensemble of possible future realizations of $\Delta A$ as a function of the total GW catalog size $N$.

In Fig.~\ref{fig:Forecast} we show the results of conducting this analysis for $\Delta\mu_{0}$ and $\Delta\sigma_{0}$ taken from the joint analysis (yellow), and also the less-meaningful memory-only analysis (blue). In large agreement with Refs.~\cite{Hubner:2019sly,Boersma:2020gxx,Hubner:2021amk,Grant:2022bla,Cheung:2024zow}, we observe that we need $\mathcal{O}(\text{\NForecastCrossing})$ events to constrain the 68\% half-width on the memory enhancement factor to be away from $0$.  Based on the horizontal axes on the top of the figure which show the time required to reach the number of observations shown on the lower horizontal axis for the event rates of \OFourRates and \OFiveRates per year for O4 and O5a~\cite{IGWN:ObservingCapabilities:2026}, this means that we may be able to observe memory by the end of the O5 observing run, since this is expected to have an observing window of $\sim$3 years, though the end of the O6 observing run is likely a more realistic timescale.

\section{Discussion}
\label{sec:discussion}

Gravitational wave memory is a key prediction of GR that is a natural target for ground-based GW detectors. It is of immense theoretical interest due to its connection to asymptotic symmetries and soft theorems and for its role in formulating a consistent theory of quantum gravity. Unfortunately, due to its low-frequency nature, it is an incredibly challenging effect to observe in individual events with the current LVK detectors. However, there is reason to believe that with enough observations, we can observe memory in a population of GW detections.

In this study, we used hierarchical Bayesian inference to measure the memory enhancement factor---the constant appearing in front of the contribution to the strain from the supermomentum flux, which is 1 in GR---across the population of GW events up to and including the recent GWTC-5.0 catalog. Unsurprisingly, we found that we can constrain this factor to \Aconstraint, which does not yield a confident measurement of the memory effect. Nonetheless, this work provides the first attempt to measure memory using hierarchical inference with a flexible model for the memory enhancement factor and thus the first analysis to successfully account for potential changes in the inferred BBH population due to memory. More generally, we also showed how to constrain the memory enhancement factor using future GWTCs and any generic waveform model, without sampling $A$, making this analysis very efficient. In large agreement with previous works, we find that we will likely need $\sim$\NForecastCrossing BBH detections to constrain the memory enhancement factor away from zero at $1\sigma$. However, since this number is expected to possibly be reached by the fifth observing run of the LVK detectors, but certainly by the sixth observing run, we find that it is likely that we will measure the GW memory effect for the first time in a population within five to ten years.

The code used for this analysis and the data that were produced are available on Zenodo via Refs.~\cite{memory_code,memory_data} under an open source license. This includes our estimates of the memory-augmented posteriors for each event.

\section*{Acknowledgments}

The authors thank Carl-Johan Haster and Colm Talbot for discussions about this work before it was carried out.
The authors also thank Jann Zosso for some insightful discussions about possible detection criteria for memory. And, finally, the authors would also like to thank the Simons Collaboration on Black Holes and Strong Gravity for fostering an environment where discussions regarding this work occurred.

K.M. is supported by NASA through the NASA Hubble Fellowship grant \#HST-HF2-51562.001-A awarded by the Space Telescope Science Institute, which is operated by the Association of Universities for Research in Astronomy, Incorporated, under NASA contract NAS5-26555. The Flatiron Institute is a division of the Simons Foundation.

This material is based upon work supported by NSF’s LIGO Laboratory which is a major facility fully funded by the National Science Foundation. This research has made use of data or software obtained from the Gravitational Wave Open Science Center (gwosc.org), a service of the LIGO Scientific Collaboration, the Virgo Collaboration, and KAGRA. This paper carries LIGO document number P2600229.

\emph{Software}: \textsc{Bilby}~\cite{bilby_paper}, \textsc{LALSuite}~\cite{2020ascl.soft12021L}, \textsc{PESummary}~\cite{Hoy:2020vys}, \textsc{GWPy}~\cite{MACLEOD2021100657}, \textsc{spherical\_functions}~\cite{mike_boyle_2023_7960723}, \textsc{NumPyro}~\cite{bingham2019pyro,Phan:2019NumPyro}, \textsc{JAX}~\cite{jax2018github}, \textsc{ArviZ}~\cite{Martin2026}, \textsc{NumPy}~\cite{numpy}, \textsc{SciPy}~\cite{2020SciPy-NMeth}, \textsc{Astropy}~\cite{astropy:2013,astropy:2018,astropy:2022}, \textsc{h5py}~\cite{the_hdf_group_2026_19194969}, \textsc{Matplotlib}~\cite{Hunter:2007}, \textsc{pandas}~\cite{reback2020pandas}.

\appendix

\section{Population model and priors}
\label{sec:appendix-popmodel}

Here we provide more details and explicit expressions for the hierarchical population model used in Sec.~\ref{sec:methodology}.\footnote{This population model follows the GWTC-5.0 Default BBH model, which is also the fiducial GWTC-4.0 Default BBH model~\cite{LIGOScientific:2025pvj,LIGOScientific:2026ctl}.}
The starting point is the likelihood in Eq.~\eqref{eq:hierarchicalLLfinal} which contains a factor $p(\lambda \mid \Lambda_\lambda)$ encoding our population model for the astrophysical parameters 
\begin{equation}
	\lambda = \left\{ m_1, q, z, \chi_1, \chi_2, \cos\theta_1, \cos \theta_2 \right\},
\end{equation}
for the primary mass $m_1$, the mass ratio $q = m_2/m_1$, the redshift $z$, and the spin magnitudes $\chi_{1/2}$ and tilts with respect to the orbital angular momentum $\theta_{1/2}$; the spin angles in the orbital plane $\phi_{1/2}$ are taken to be azimuthally uniform, while the source locations and the inclinations are taken to follow isotropy, so that neither appears explicitly in our population model.

We impose a population model on the BBH parameters $\lambda$ that factorizes as
\begin{align}
\label{eq:appendix-factorization}
p(\lambda \mid \Lambda_\lambda)
&= p(m_{1} \mid \Lambda_{m_{1}})\, p(q \mid m_{1}, \beta)\, p(z \mid \lambda_{z}) \\
&\quad \times p(\cos \theta_{1}, \cos \theta_{2} \mid \Lambda_{\theta})\,
p(\chi_{1} \mid \Lambda_{\chi})\, p(\chi_{2} \mid \Lambda_{\chi}), \nonumber
\end{align}
with the two component spin azimuths $\phi_{1}, \phi_{2}$ taken to be independent and uniform on $[0, 2\pi]$, thus contributing a mere overall constant factor of $(2\pi)^{-2}$ to the joint density. The astrophysical hyperparameters are
\begin{align}
\Lambda_{\lambda} = \big\{&\Lambda_{m_{1}},\, \beta,\, \lambda_{z},\, \Lambda_\theta,\, \Lambda_{\chi}\big\},
\end{align}
for which $\Lambda_{m_{1}} = \{\alpha_{1}, \alpha_{2}, b, \mu_{1}, \sigma_{1}, \mu_{2}, \sigma_{2}, f_{\mathrm{bpl}}, f_{1}, f_{2}\}$, $\Lambda_\theta = \{f_{\mathrm{iso}}, \mu_{\theta}, \sigma_{\theta}\}$, and $\Lambda_{\chi} = \{\mu_{\chi}, \sigma_{\chi}\}$.

Factors of the astrophysical model are described below, while
the memory enhancement factor $A$ of this work is modeled as a Gaussian with uniform hyperpriors on its mean and width, as stated in Sec.~\ref{sec:hierarchicalinference} of the main text. 
All hyperparameters are summarized in Table~\ref{tab:priors}.

\begin{table}[t]
	\caption{Hyperpriors that are used in the population analysis. All hyperparameters carry independent uniform priors except for the mass-mixture weights, which are instead drawn from a flat Dirichlet. The bounds on the memory hyperparameters $\mu_{\Lambda}, \sigma_{\Lambda}$ are auto-scaled by $\hat{A}_{\max} \equiv \max_{i, s} |\hat{A}_{i, s}|$, the largest sample-level memory mean across all events. Mass quantities are in $M_{\odot}$; all other quantities are dimensionless.}
	\label{tab:priors}
	\renewcommand{\arraystretch}{1.15}
	\begin{ruledtabular}
	\begin{tabular}{lll}
	Parameter & Distribution & Range \\
	\hline
	\multicolumn{3}{l}{\textit{Primary mass}} \\
	$\alpha_{1}$, $\alpha_{2}$ & Uniform & $[-4, 12]$ \\
	$m_{b}$ (via $b$) & Uniform & $[20, 50]\,M_{\odot}$ \\
	$(f_{\mathrm{bpl}}, f_{1}, f_{2})$ & Dirichlet & $\mathrm{Dir}(1, 1, 1)$ \\
	$\mu_{1}$ & Uniform & $[5, 20]\,M_{\odot}$ \\
	$\sigma_{1}$ & Uniform & $[0.05, 6]\,M_{\odot}$ \\
	$\mu_{2}$ & Uniform & $[25, 60]\,M_{\odot}$ \\
	$\sigma_{2}$ & Uniform & $[0.05, 10]\,M_{\odot}$ \\[2pt]
	\hline
	\multicolumn{3}{l}{\textit{Mass ratio}} \\
	$\beta$ & Uniform & $[-2, 7]$ \\[2pt]
	\hline
	\multicolumn{3}{l}{\textit{Redshift}} \\
	$\lambda_{z}$ & Uniform & $[-10, 10]$ \\[2pt]
	\hline
	\multicolumn{3}{l}{\textit{Spin tilts}} \\
	$f_{\mathrm{iso}}$ & Uniform & $[0, 1]$ \\
	$\mu_{\theta}$ & Uniform & $[-1, 1]$ \\
	$\sigma_{\theta}$ & Uniform & $[0.05, 10]$ \\[2pt]
	\hline
	\multicolumn{3}{l}{\textit{Spin magnitudes}} \\
	$\mu_{\chi}$ & Uniform & $[0, 0.7]$ \\
	$\sigma_{\chi}$ & Uniform & $[0.01, 1]$ \\[2pt]
	\hline
	\multicolumn{3}{l}{\textit{Memory}} \\
	$\mu_{\Lambda}$ & Uniform & $[-1.5\,\hat{A}_{\max},\, 1.5\,\hat{A}_{\max}]$ \\
	$\sigma_{\Lambda}$ & Uniform & $[0,\, 1.5\,\hat{A}_{\max}]$ \\
	\end{tabular}
	\end{ruledtabular}
	\end{table}	

\subsection{Primary mass}
\label{sec:appendix-m1}

The primary mass is supported on $[m_{\min}, m_{\max}] = [3, 300]\, M_{\odot}$ and described by a three-component mixture of a continuous broken power law and two truncated Gaussian peaks,
\begin{align}
\label{eq:m1-mixture}
p(m_{1} \mid \Lambda_{m_{1}})
&= f_{\mathrm{bpl}}\, p_{\mathrm{bpl}}(m_{1} \mid \alpha_{1}, \alpha_{2}, m_{b}) \nonumber \\
&\quad + f_{1}\, \mathcal{N}_{[m_{\min}, m_{\max}]}(m_{1} \mid \mu_{1}, \sigma_{1}) \nonumber \\
&\quad + f_{2}\, \mathcal{N}_{[m_{\min}, m_{\max}]}(m_{1} \mid \mu_{2}, \sigma_{2}),
\end{align}
where $\mathcal{N}_{[a, b]}(x \mid \mu, \sigma)$ denotes a Gaussian with mean $\mu$ and width $\sigma$ truncated and renormalized to $[a, b]$, and the mixture weights $(f_{\mathrm{bpl}}, f_{1}, f_{2})$ are drawn from a flat Dirichlet distribution, $(f_{\mathrm{bpl}}, f_{1}, f_{2}) \sim \mathrm{Dir}(1, 1, 1)$, so they are positive and sum to unity. The broken power law is piecewise on $[m_{\min}, m_{\max}]$ and continuous at the break,
\begin{align}
\label{eq:bpl}
p_{\mathrm{bpl}}(m_{1}) \propto
\begin{cases}
m_{1}^{-\alpha_{1}} & m_{\min} \le m_{1} < m_{b}, \\[2pt]
m_{b}^{\alpha_{2} - \alpha_{1}}\, m_{1}^{-\alpha_{2}} & m_{b} \le m_{1} \le m_{\max},
\end{cases}
\end{align}
normalized to unity over $[m_{\min}, m_{\max}]$. The break mass is reparametrized as
\begin{align}
m_{b} = m_{\min} + b\, (m_{\max} - m_{\min}),
\end{align}
restricted by the prior on $b$ to $m_{b} \in [20, 50]\, M_{\odot}$.

\subsection{Mass ratio}
\label{sec:appendix-q}

The mass ratio $q \equiv m_{2}/m_{1}\leq1$ follows a power law conditioned on the primary mass,
\begin{align}
\label{eq:q-density}
p(q \mid m_{1}, \beta) \propto q^{\beta} \quad \text{on}\quad\ q \in [m_{\min}/m_{1},\, 1],
\end{align}
with the lower edge enforcing $m_{2} \ge m_{\min} = 3\, M_{\odot}$.

\subsection{Redshift}
\label{sec:appendix-z}

The redshift distribution scales as a power law in $1+z$ that is multiplied by the differential comoving volume per source-frame time,
\begin{align}
\label{eq:redshift-density}
p(z \mid \lambda_{z}) \propto (1 + z)^{\lambda_{z} - 1}\, \frac{\mathrm{d} V_{c}}{\mathrm{d} z},
\end{align}
where the factor of $(1+z)^{-1}$ converts the detector-frame rate of mergers to the source frame. For this we use a flat $\Lambda$CDM cosmology with $H_{0} = 67.90$~km/s/Mpc and $\Omega_{m} = 0.3065$, consistent with Planck~2015~\cite{Planck:2015fie}, and restrict the support to $z \in [0, 2.5]$.

\subsection{Spin tilts}
\label{sec:appendix-tilt}

The cosines of the spin tilt angles, $\cos \theta_{1}, \cos \theta_{2} \in [-1, 1]$, are jointly modeled as a mixture between an isotropic component uniform on $[-1, 1]^{2}$ with density $1/4$, and a product of $[-1, 1]$-truncated Gaussians with shared mean $\mu_{\theta}$ and width $\sigma_{\theta}$,
\begin{align}
\label{eq:tilt-density}
p(\cos \theta_{1}, \cos \theta_{2} \mid \Lambda_\theta) =\;& \frac{f_{\mathrm{iso}}}{4}  \\
&\hspace{-30pt} + (1 - f_{\mathrm{iso}}) \prod_{i=1}^{2} \mathcal{N}_{[-1, 1]}(\cos \theta_{i} \mid \mu_{\theta}, \sigma_{\theta}). \nonumber
\end{align}
This is the ``non-independently but identically distributed'' prescription of the GWTC-5.0 \textsc{Gaussian Component Spins} model~\cite{LIGOScientific:2026ctl}. The isotropic-mixture weight $f_{\mathrm{iso}}$ allows for a sub-population of randomly oriented spins, while the shared $(\mu_{\theta}, \sigma_{\theta})$ tie the two component-spin orientations to a common preferred-alignment distribution.

\subsection{Spin magnitudes}
\label{sec:appendix-spin}

The two dimensionless spin magnitudes $\chi_{i}$ are modeled using independent and identically distributed truncated Gaussians on $[0, 1]$,
\begin{align}
\label{eq:spin-mag-density}
p(\chi_{i} \mid \mu_{\chi}, \sigma_{\chi}) = \mathcal{N}_{[0, 1]}(\chi_{i} \mid \mu_{\chi}, \sigma_{\chi}) \, , \quad i \in \{1, 2\},
\end{align}
with shared mean and width across the two components.

\subsection{Selection function}
\label{sec:appendix-memory-selection}

Possible selection effects are accounted for through the detection efficiency $\xi(\Lambda)$ in Eq.~\eqref{eq:poplike}, which is evaluated as a Monte-Carlo average over the LVK Collaboration's cumulative O1--O4b simulated-injection campaign~\cite{Essick:2025zed,LVK:GWTC5CumulativeSensitivity}. Both this Monte-Carlo sum and the per-event $N$-sample sums approximating $\mathcal{L}_{i}(\Lambda)$ in Eq.~\eqref{eq:hierarchicalLLfinal} are combined with smooth penalties that suppress certain regions of the hyperparameter space where the corresponding effective sample sizes become too small for the estimator to be reliable~\cite{Tiwari:2017ndi,Farr:2019rap,Essick:2022ojx,Talbot:2023pex,Heinzel:2025ogf}. Writing $w_{i,s} \equiv p(\lambda_{i,s} \mid \Lambda_{\lambda})/p_{\mathrm{pe}}(\lambda_{i,s})$ for the importance weight of the $s$-th posterior sample of event $i$ relative to its parameter-estimation prior $p_{\mathrm{pe}}$, and $w_{j}^{\mathrm{sel}} \equiv p(\lambda_{j} \mid \Lambda_{\lambda})/p_{\mathrm{inj}}(\lambda_{j})$ for the analogous weight on the $j$-th of the $N_{\mathrm{draw}}$ recovered injections drawn from the injection distribution $p_{\mathrm{inj}}$, the per-event and selection effective sample sizes are
\begin{align}
\label{eq:neff-event}
N_{\mathrm{eff}, i}(\Lambda) &\equiv \frac{\left(\sum_{s} w_{i,s}\right)^{2}}{\sum_{s} w_{i,s}^{2}},\\
\label{eq:neff-sel}
N_{\mathrm{eff,sel}}(\Lambda) &\equiv \frac{\left(\sum_{j} w_{j}^{\mathrm{sel}}\right)^{2}}{\sum_{j} \left(w_{j}^{\mathrm{sel}}\right)^{2} - \left(\sum_{j} w_{j}^{\mathrm{sel}}\right)^{2} \! / N_{\mathrm{draw}}}.
\end{align}
We add to the log-likelihood a smooth penalty for each $N_{\mathrm{eff}}$ that ramps linearly below a threshold $N_{\mathrm{crit}}$,
\begin{align}
\label{eq:neff-penalty}
\log \Pi(N_{\mathrm{eff}} \mid N_{\mathrm{crit}}) = \min\!\left(0,\; \frac{N_{\mathrm{eff}} - N_{\mathrm{crit}}}{0.05\, N_{\mathrm{crit}}}\right),
\end{align}
applied with $N_{\mathrm{crit}} = N_{\mathrm{obs}}$ to $\min_{i}\, N_{\mathrm{eff}, i}(\Lambda)$ and with $N_{\mathrm{crit}} = 4\, N_{\mathrm{obs}}$ to $N_{\mathrm{eff,sel}}(\Lambda)$.

\bibliography{refs}

\end{document}